# Fermi surface reconstruction in high-$T_c$ superconductors

**Louis Taillefer**

Canadian Institute for Advanced Research
Regroupement Québécois sur les Matériaux de Pointe
Département de physique, Université de Sherbrooke, Sherbrooke, Canada

E-mail: Louis.Taillefer@USherbrooke.ca

**Abstract**. The recent observation of quantum oscillations in underdoped high-$T_c$ superconductors, combined with their negative Hall coefficient at low temperature, reveals that the Fermi surface of hole-doped cuprates includes a small electron pocket. This strongly suggests that the large hole Fermi surface characteristic of the overdoped regime undergoes a reconstruction caused by the onset of some order which breaks translational symmetry. Here we consider the possibility that this order is "stripe" order, a form of combined charge / spin modulation observed most clearly in materials like Eu-doped and Nd-doped LSCO. In these materials, the onset of stripe order coincides with major changes in transport properties, providing strong evidence that stripe order is indeed the cause of Fermi-surface reconstruction. We identify the critical doping where this reconstruction occurs and show that the temperature dependence of transport coefficients at that doping is typical of metals at a quantum critical point. We discuss how the pseudogap phase may be a fluctuating precursor of the stripe-ordered phase.

## 1. Phase diagram

The doping phase diagram of hole-doped cuprates is sketched in Fig. 1a. With increased doping $p$, the materials go from being antiferromagnetic insulators at zero doping to more or less conventional metals at high doping. The overdoped metallic state is characterized by a single large hole Fermi surface whose volume contains $1 + p$ holes per Cu atom, as determined by angle-dependent magneto-resistance (ADMR) [1] and angle-resolved photoemission spectroscopy (ARPES) [2]. The low-temperature Hall coefficient $R_H$ is positive and equal to $V / e (1 + p)$ [3], as expected for a single-band metal with a hole density $n = 1 + p$. The electrical resistivity $\rho(T)$ exhibits the standard $T^2$ temperature dependence of a Fermi liquid [4].

At intermediate doping, between the insulator and the metal, there is a central region of superconductivity, delineated by a critical temperature $T_c$ which can rise to values of order 100 K. Above the maximal $T_c$, near optimal doping, the normal state is a "strange metal", characterized by a resistivity which is linear in temperature instead of quadratic. In the midst of this strange-metal region, the enigmatic "pseudogap phase" sets in, below a crossover temperature $T^*$ where most physical properties undergo a smooth yet significant change [5].

Elucidating the nature of the pseudogap phase is key to understanding high-temperature superconductivity. Two main scenarios have been proposed [6]: fluctuating superconductivity – a precursor to the long-range coherence which sets in below $T_c$ – versus some other ordered state. For hole-doped cuprates, a number of different types of order have been proposed, including "stripe order"

[7], *d*-density-wave order [8] and orbital currents [9]. In this article, we review some recent transport measurements performed in magnetic fields high enough to suppress superconductivity and thus give access to the normal-state behaviour of hole-doped cuprates down to low temperature. These shed new light on the ground state of the pseudogap phase, raising hopes of finally unravelling this mysterious phenomenon.

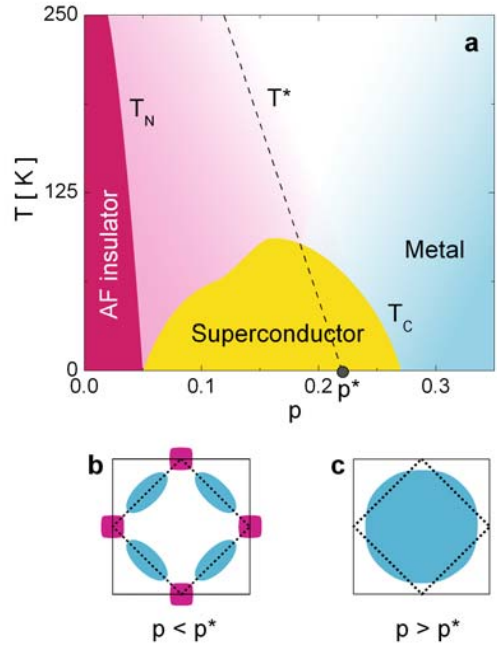

**Figure 1. Phase diagram of hole-doped high-$T_c$ superconductors.**

**a)** Schematic doping dependence of the antiferromagnetic ($T_N$) and superconducting ($T_c$) transition temperatures and the pseudogap crossover temperature $T^*$. The fact that the large hole-like Fermi surface characteristic of the overdoped metallic state, sketched in panel (**c**), is modified in the underdoped region (see text) implies that there is a critical doping $p^*$ where Fermi-surface reconstruction occurs. **b)** Schematic drawing of one possible reconstruction, that would result from an order with $(\pi, \pi)$ wavevector, as in the antiferromagnetic state.

## 2. Quantum oscillations

In 2007, quantum oscillations were finally observed in a high-$T_c$ superconductor [10]. A key factor in the ability to detect these oscillations, whose amplitude decreases exponentially with increased disorder, was the high degree of ortho-II oxygen order in single crystals of $YBa_2Cu_3O_y$ (YBCO) [11]. Quantum oscillations result from the Landau quantization of states in a magnetic field and the orbiting motion of quasiparticles around the various pockets of the Fermi surface in a metal. Their very observation confirms the existence of a coherent closed Fermi surface and their frequency $F$ is a direct measure of the Fermi surface area, via the relation $F = n\Phi_0$, where $n$ is the carrier density enclosed by the particular Fermi surface associated with a given frequency, and $\Phi_0$ is the quantum of flux.

First observed in the electrical resistance (both Hall and longitudinal; the Shubnikov-de Haas effect) [10], the same oscillations were soon also detected in the de Haas-van Alphen effect (magnetization) [12]. The Fourier transform of the oscillatory spectrum in $YBa_2Cu_3O_{6.5}$, reproduced in Fig. 2 (from [13]), reveals a single frequency at $F = 540$ T [10, 12]. In 2008, quantum oscillations were observed in strongly overdoped $Tl_2Ba_2CuO_{6+\delta}$ (Tl-2201), which also reveal a single frequency, but now at $F = 18$ kT [14] (see Fig. 2). This large value matches the area derived previously from ADMR and ARPES measurements on the same material at a similar doping and agrees with $n = 1 + p$.

The contrast between Tl-2201 at $p \approx 0.25$ and YBCO at $p = 0.1$ is dramatic: the Fermi surface area differs by a factor 30 (see Fig. 2). Note that the small pockets detected in underdoped YBCO are not a special feature of the band structure of that particular material, since similar quantum oscillations were observed in the stoichiometric underdoped cuprate $YBa_2Cu_4O_8$ [15, 16], whose band structure is significantly different [17]. Therefore, this transformation of the Fermi surface from large cylinder to small pockets is a robust signature of the pseudogap phase, which must occur at a $T = 0$ critical doping $p^*$ somewhere between 0.1 and 0.25 (see Fig. 1).

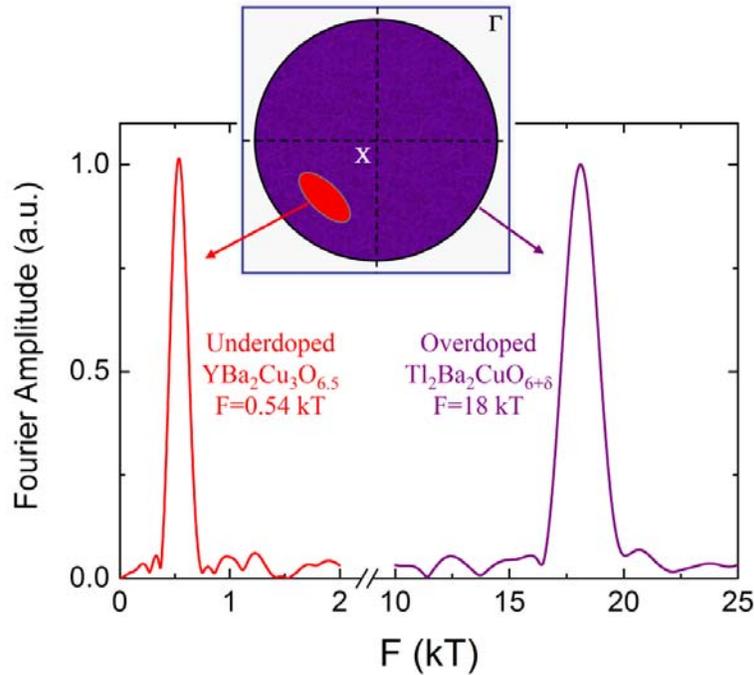

**Figure 2. Quantum oscillations.**

Fourier transform of the quantum oscillations detected in YBCO at $p = 0.1$ and Tl-2201 at $p = 0.25$. Each reveals a single frequency, but with vastly different values, as indicated. This shows that the Fermi surface in the underdoped regime includes a pocket which is much smaller than that of the overdoped regime, as sketched in the inset. Courtesy of Cyril Proust [13].

## 3. Electron Fermi surface

A second important fact is that the low-frequency oscillations in $YBa_2Cu_3O_{6.5}$ and $YBa_2Cu_4O_8$ are observed on the background of a negative Hall coefficient $R_H$ at low temperature [18] (Fig. 3a). As a function of temperature, $R_H(T)$ goes from small and positive at high temperature to large and negative as $T \to 0$ (Fig. 3b). Given that $R_H \sim 1/n$ and $F \sim n$, this is consistent with the transition from large to small Fermi surface revealed by quantum oscillations, but the fact that $R_H(T \to 0) < 0$ implies that the small Fermi surface seen in the underdoped regime must in fact be an electron-like pocket.

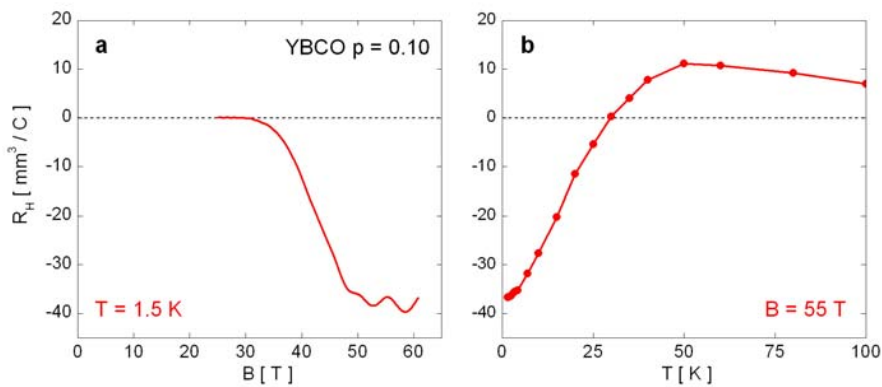

**Figure 3. Hall coefficient in YBCO.**

Hall coefficient in YBCO at $p = 0.1$, as a function of magnetic field at $T = 1.5$ K (**a**) and as a function of temperature at $B = 55$ T (**b**). The fact that quantum oscillations are observed on a large negative background implies that they arise from orbits around a closed electron-like Fermi surface pocket. From [18].

The emergence of an electron pocket in the Fermi surface of these hole-doped materials is of fundamental significance: it immediately suggests that the transformation of the Fermi surface is caused by the onset of a new periodicity, typically imposed by some density-wave order [19]. The simplest case to visualize is commensurate $(\pi, \pi)$ antiferromagnetic order, which would cause the large hole-like Fermi surface of cuprates to be reconstructed into small hole and electron pockets [20], located respectively at $(\pi/2, \pi/2)$ and $(\pi, 0)$, as sketched in Fig. 1b. Because $d$-density-wave order breaks translational symmetry in the same way, a similar reconstruction is produced [21]. However, a different reconstruction is expected for "stripe order", a state with both charge and spin modulations, with wavevectors $(\pi, \pi+\varepsilon)$ and $(\pi, \pi+2\varepsilon)$, respectively [22]. In the case of commensurate anti-phase stripe order ($\varepsilon = 1/8$), the Fermi surface is generically predicted to have hole pockets, electron pockets and quasi-1D open sheets [23], as sketched in Fig. 4. Note that quantum oscillations do not allow us to locate the position of the associated Fermi pockets in $k$-space, so the observed electron pocket can in principle be anywhere in the (reconstructed) Brillouin zone, and the three types of order just mentioned are *a priori* consistent with the evidence so far.

The Hall coefficient of YBCO at $p = 0.12$ is shown in Fig. 5b (from [18]). We see that $R_H(T)$ starts to drop below 100 K and changes sign at $T_0 = 70$ K [18]. We emphasize that this drop cannot be caused by a vortex (flux flow) contribution to the Hall effect because it is entirely independent of magnetic field, for fields ranging all the way from $B \approx 0$ to $B = 45$ T. In other words, $T_0$ is constant while the superconducting transition goes from $T_c = 66$ K at $B = 0$ to $T_c \approx 0$ at $B = 45$ T (see Supplementary Information in [18]), and is therefore a property of the normal state. This independence of $T_0$ on field also shows that the modification of the Fermi surface implied by the sign change in $R_H$ is characteristic of the zero-field pseudogap phase, not some field-induced ordered state.

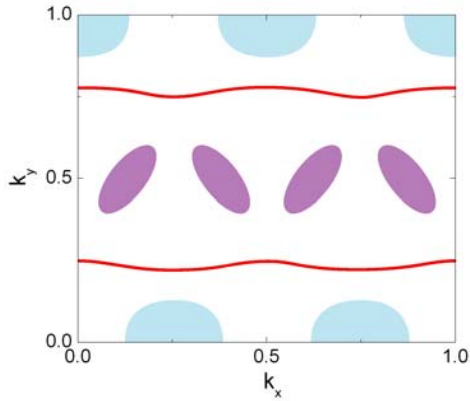

**Figure 4. Fermi surface in stripe phase.**

Calculated Fermi surface in the anti-phase stripe-ordered state of a hole-doped cuprate at $p = 1/8$, for a finite spin potential. Three types of Fermi surfaces are generically predicted: electron pockets (pale blue), hole pockets (pink) and quasi-1D open sheets (red lines). From [23].

A drop in the normal-state $R_H(T)$ is a generic feature of hole-doped cuprates near $p = 1/8$, observed in $Bi_2La_{2-x}Ba_xCuO_{6+\delta}$ [24], $La_{2-x}Sr_xCuO_4$ (LSCO) [25], $La_{2-x}Ba_xCuO_4$ [26], Nd-doped LSCO (Nd-LSCO) [27] and Eu-doped LSCO (Eu-LSCO) [28], in addition to $YBa_2Cu_3O_y$ and $YBa_2Cu_4O_8$ [18]. The depth of the drop in a particular sample will depend on the relative mobility of electron-like and hole-like carriers. In Fig. 5b, we compare YBCO and Eu-LSCO [29] at $p = 1/8$. The drops in $R_H(T)$ are so strikingly similar, it would be surprising if the underlying mechanism were not the same. Now, in Eu-LSCO, the drop in $R_H(T)$ coincides with the onset of charge "stripe" order measured by resonant soft X-ray diffraction [30], as reproduced in Fig. 5a. This is compelling evidence that stripe order causes a Fermi-surface reconstruction which shows up as a pronounced change in the Hall coefficient.

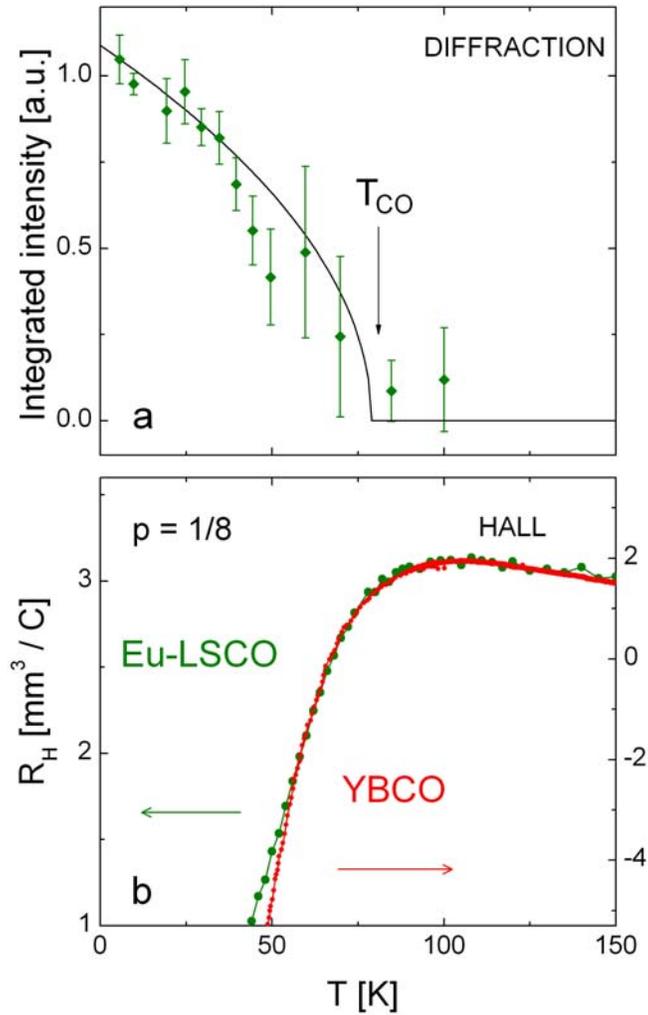

**Figure 5. Stripe order and Hall coefficient in Eu-LSCO at $p = 1/8$.**

**a)** Temperature dependence of charge order in Eu-LSCO at $p = 1/8$, as detected by resonant soft X-ray diffraction (from [30]). **b)** Hall coefficient vs temperature measured in $B = 15$ T for Eu-LSCO (green; left axis; from [29]) and YBCO (red; right axis; from [18]) at $p = 1/8$.

## 4. Stripe order

In order to further investigate the impact of stripe order on the Fermi surface of cuprates, we have studied Nd-LSCO, a material isostructural to Eu-LSCO, which exhibits very similar charge and spin ordering. The onset of charge order in the two materials occurs at essentially the same temperature, $T_{CO}$, as a function of doping. Two measures of $T_{CO}$, obtained respectively from X-ray diffraction [30, 31] and nuclear quadrupole resonance (NQR) [32], are plotted in the phase diagram of Fig. 6. In particular, we have compared two samples of Nd-LSCO, respectively at $p = 0.20$ and $p = 0.24$ [33]. In Fig. 7, we show the in-plane resistivity $\rho(T)$ and Hall coefficient $R_H(T)$, from [33], and the Seebeck coefficient (or thermopower) $S$, plotted as $S / T$, from [34], as a function of temperature.

At $p = 0.24$, $R_H(T)$ is flat at low temperature (see Fig. 7c) and equal to the value expected of a single large cylinder containing $1 + p$ holes, namely $R_H = + V / e (1 + p)$, just as found in Tl-2201 at a similar doping [3]. The other two coefficients, $\rho(T)$ and $S / T$, are equally monotonic and featureless. By contrast, at $p = 0.20$, all three transport coefficients exhibit a pronounced upturn below 40 K, which shows that the Fermi surface has undergone a significant modification. These simultaneous upturns coincide with the onset of charge order detected by NQR at $T_{CO} = 40$ K [30], as reproduced in Fig. 7a. So as in the case of Eu-LSCO at $p = 1/8$, there is little doubt that stripe order causes Fermi-surface reconstruction in Nd-LSCO.

An intriguing difference is that $R_H(T)$ rises below $T_{CO}$ at $p = 0.20$ (Fig. 7c), while it drops at $p = 1/8$ (Fig. 5b). If the drop in $R_H(T)$ near $p = 1/8$ is caused by a high-mobility electron pocket, then the rise at $p = 0.20$ suggests that this electron pocket is absent at higher doping. Calculations of the Hall coefficient in the stripe-ordered phase [35] reveal that a negative $R_H$ requires a finite spin-stripe potential, the cause of a robust electron pocket in the Fermi surface [23] (as in Fig. 4). The observed evolution from a rise in $R_H$ just below $p^*$ to a drop in $R_H$ (in some cases to negative values) near $p = 1/8$, may therefore reflect an increase in the spin-stripe potential relative to the charge-stripe potential. This would seem consistent with the fact that static spin order detected by muon spin relaxation, whose onset at $T_M$ is plotted in Fig. 6, is absent at $p = 0.20$ and strongest at $p = 0.12$ [36].

## 5. The pseudogap phase

What relation might there be between the stripe-ordered phase which sets in below $T_{CO}$ and the mysterious pseudogap phase delineated by the higher crossover temperature $T^*$, sketched in Fig. 1? One way to define $T^*$ is through the resistivity [5], as the temperature below which $\rho(T)$ deviates from its linear dependence at high temperature. A resistively-defined $T^*$, which we label $T_\rho^*$, was first reported for YBCO [37], where the deviation is downwards. In LSCO, however, the deviation is upwards [38], as indeed in Nd-LSCO [27, 33, 39]. The difference may simply reflect two limits: the clean limit, relevant for YBCO, where the loss of inelastic scattering caused by the opening of the pseudogap is more important than the loss in carrier density, and the dirty limit, relevant to LSCO and Nd-LSCO, where the reverse is true [33]. In Fig. 6, we plot $T_\rho^*$ vs $p$ for Nd-LSCO (from [33]). Both $T_{CO}$ and $T_\rho^*$ appear to end at the same critical point $p^* \approx 0.24$, and $T_\rho^* \approx 2\,T_{CO}$. The onset of spin modulation seen in neutron diffraction (dashed line in Fig. 6) also appears to end at $p^*$ [39].

In this context, it seems natural to interpret the pseudogap phase as a fluctuating precursor of the long-range ordered state that sets in at lower temperature [40]. The onset of fluctuations at a temperature $T^*$ well above $T_{CO}$ may be understood from calculations which show quite generally that, for layered materials, precursors of the ordered state appear when the correlation length of the fluctuating order parameter exceeds the thermal de Broglie wavelength of the charge carriers [41]. Precursor features include a pseudogap and hot spots on the Fermi surface. Quantitative agreement between calculations on the Hubbard model at moderate coupling [42] and neutron diffraction measurements [43] confirm this interpretation in the case of electron-doped high-$T_c$ superconductors.

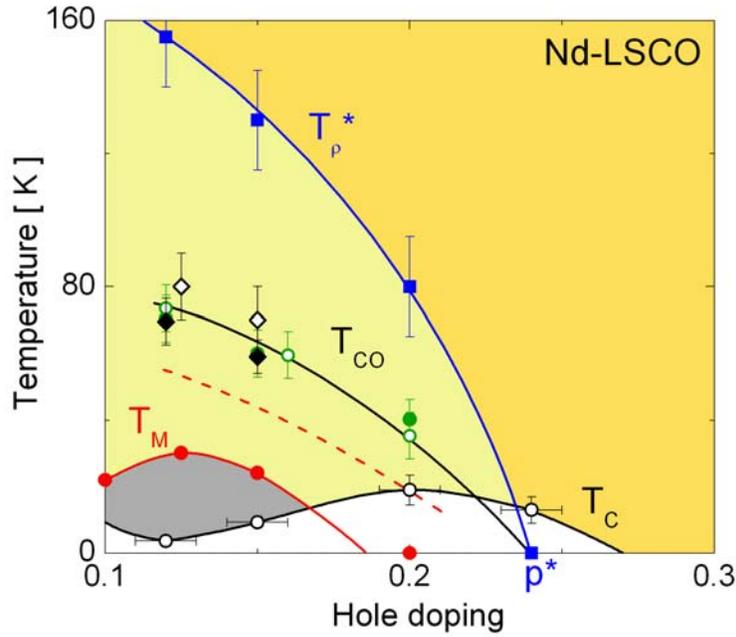

**Figure 6. Phase diagram of Nd-LSCO.**

Temperature-doping phase diagram of Nd-LSCO showing the superconducting phase below $T_c$ [open black circles] and the pseudogap region delineated by the crossover temperature $T_\rho^*$ [blue squares]. Also shown is the region where static magnetism is observed below $T_M$ [red circles] and charge order is detected below $T_{CO}$ [black diamonds and green circles]. These onset temperatures are respectively defined as the temperature below which: 1) the resistance is zero; 2) the in-plane resistivity $\rho(T)$ deviates from its linear dependence at high temperature; 3) an internal magnetic field is detected by zero-field muon spin relaxation (μSR); 4) charge order is detected by either X-ray diffraction or NQR. All lines are a guide to the eye. The red dashed line shows the onset of spin modulation as detected by neutron diffraction [39]. The blue line above $p = 0.20$ is made to end at $p = 0.24$, thereby defining the critical doping where $T_\rho^*$ goes to zero as $p^* = 0.24$. Experimentally, this point must lie in the range $0.20 < p^* \leq 0.24$, since $\rho(T)$ remains linear down to the lowest temperature at $p = 0.24$. $T_M$ is obtained from the μSR measurements of [36]. The red line is made to end below $p = 0.20$, as no static magnetism was detected at $p = 0.20$ down to $T = 2$ K. $T_{CO}$ is obtained from hard X-ray diffraction on Nd-LSCO (full black diamonds [31]) and from resonant soft X-ray diffraction on Eu-LSCO (open diamonds [30]). The onset of charge order has been found to coincide with the wipeout anomaly in NQR, reproduced here from [32] for Nd-LSCO (closed green circles) and Eu-LSCO (open green circles).

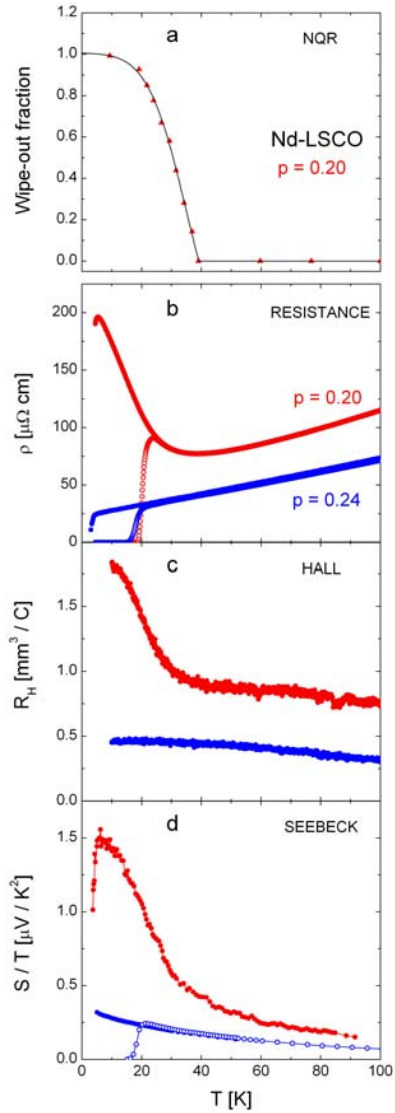

**Figure 7. Stripe order and transport coefficients in Nd-LSCO.**

**a)** Charge ordering in Nd-LSCO at $p = 0.20$, as measured by the loss of NQR intensity (from [32]). At dopings $p = 0.12$ and $p = 0.15$, where both X-ray diffraction and NQR where measured on Nd-LSCO, the lost (or "wipe-out") fraction of the intensity present at 100 K tracks the increase in the intensity of superlattice peaks detected with X-rays. At $p = 0.20$, the onset of charge order is $T_{CO} = 40 \pm 6$ K [32]. *Lower panels*: transport coefficients in two samples of Nd-LSCO, respectively with $p = 0.20$ (red) and at $p = 0.24$ (blue): **b)** in-plane electrical resistivity $\rho$ in a magnetic field $B = 0$ (open symbols) and 15 T (closed symbols) (from [33]); **c)** Hall coefficient $R_H$ in 15 T (from [33]); **d)** Seebeck coefficient $S$ plotted as $S / T$ for $B = 0$ (open symbols) and 15 T (closed symbols) (from [34]). Note how at $p = 0.20$ all coefficients show a pronounced and simultaneous upturn starting at a temperature which coincides with the onset of charge order – strong evidence for a scenario of Fermi-surface reconstruction by stripe order.

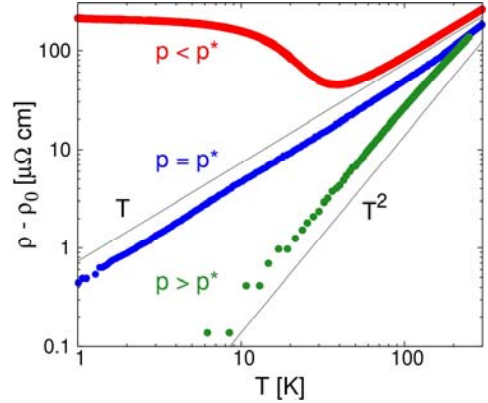

**Figure 8. Three regimes of quantum criticality.**

Temperature dependent part of the resistivity, $\rho(T) - \rho_0$, vs log$T$ for Nd-LSCO with $p = 0.20$ ($p < p^*$) and $p = 0.24$ ($p = p^*$), from [32], compared to that of LSCO with $p = 0.30$ ($p > p^*$), from [4]. $\rho_0$ is the value to which $\rho(T)$ extrapolates at $T = 0$; for Nd-LSCO at $p = 0.20$, the extrapolation is based only on data above 80 K. Taken from [34].

## 6. Quantum critical point

The Fermi-surface reconstruction in Nd-LSCO occurs between $p = 0.20$ and $p = 0.24$. The phase diagram in Fig. 6 suggests that at $T = 0$ it takes place at $p^* \approx 0.24$. Above this critical doping, the Fermi surface is in its pristine large hole-like state, and it is profoundly modified below $p^*$. While the value of $p^*$ may be somewhat different in YBCO, quantum oscillations show that there must also be a $T = 0$ critical point in that material at which Fermi-surface reconstruction occurs, somewhere above 0.1, and probably below 0.25. From an analysis of various physical properties, it has been proposed that $p^* \approx 0.19$ [44].

It is interesting to scrutinize the low-temperature properties of the metallic state at $p^*$. Close inspection reveals that $\rho(T)$ is linear down to the lowest temperature [33] and $S / T$ exhibits a perfect $\log(1 / T)$ dependence below 100 K [34]. This $\log(1 / T)$ dependence of $S / T$ is reminiscent of the $\log(1 / T)$ dependence observed in $C_e / T$, the electronic specific heat divided by temperature, at the quantum critical point of various heavy-fermion metals [45]. The similarity with the antiferromagnetic compound CeCu$_{6-x}$Au$_x$ [46], for example, is remarkable (see [34]), both materials displaying the three distinctive regimes of quantum criticality, whereby $S / T$ and $C_e / T$ are relatively flat in the Fermi-liquid state, logarithmically divergent at the critical point, and a jump in the ordered state [34].

This qualitative similarity suggests that $p^*$ in Nd-LSCO is a quantum critical point at which a quantum phase transition occurs. This is reinforced by the resistivity behavior, which also displays the three regimes characteristic of a quantum critical point, as shown in Fig. 8 (from [34]): quadratic in the Fermi-liquid state, linear at the critical point, and an upturn below that point.

There is also a strong similarity between Nd-LSCO and the electron-doped cuprate Pr$_{2-x}$Ce$_x$CuO$_4$ (PCCO), where the case for a quantum critical point is well established [47]. In the $T \rightarrow 0$ limit, both $R_H(T)$ and $S / T$ in PCCO show an abrupt change as the doping $x$ drops below the critical doping $x_c$, signaling a change in Fermi surface from a large hole cylinder to a combination of small electron and hole pockets [48, 49]. The two coefficients track each other, as equivalent measures of the effective carrier density [48]. At $x = x_c$, $\rho(T)$ is again linear in temperature at low temperature [50]. These typical signatures of a quantum critical point have been attributed to the loss of antiferromagnetic order near $x_c$ [43], and the quantum fluctuations thereof.

In a model of charge carriers on a three-dimensional Fermi surface scattered by two-dimensional antiferromagnetic spin fluctuations, transport properties near the magnetic quantum critical point are found to be dominated by "hot spots", points on the Fermi surface connected by the ordering wavevector. In this case, calculations show that $\rho(T) \sim T$, $C_e / T \sim \log(1 / T)$ and $S / T \sim \log(1 / T)$ [51]. More generally, both $\rho(T) \sim T$ and $C_e / T \sim \log(1 / T)$ follow naturally from a marginal Fermi liquid phenomenology [52].

## 7. Summary

The low frequency of quantum oscillations and the drop in Hall coefficient to deeply negative values observed in YBCO near $p = 1/8$ demonstrate that the large hole Fermi surface of overdoped cuprates undergoes a profound reconstruction in the pseudogap phase. In the case of Eu-LSCO and Nd-LSCO, this reconstruction is clearly caused by the onset of stripe order. Given the striking similarity between YBCO and Eu-LSCO in the way $R_H(T)$ drops below 100 K at $p = 1/8$, it is tempting to invoke the same mechanism in YBCO. However, the lack of evidence for static stripe order in YBCO at $p = 1/8$ raises an interesting question: are fluctuating charge / spin modulations sufficient to alter the Fermi surface? This would point to a scenario where the pseudogap phase is a fluctuating precursor of a long-range stripe order that only sets in at lower temperature [7, 40, 53]. Further experimental and theoretical investigations are needed to answer this question and explore this scenario.

**Acknowledgments**
I wish to thank my collaborators on studies of the Fermi surface of cuprates: Luis Balicas, Doug Bonn, Olivier Cyr-Choinière, Ramzy Daou, Nicolas Doiron-Leyraud, Walter Hardy, Nigel Hussey, Francis Laliberté, David LeBoeuf, Shiyan Li, Ruixing Liang, Cyril Proust, Hidenori Takagi, and Jianshi Zhou. I wish to acknowledge stimulating discussions on this topic with Kamran Behnia, Sudip Chakravarty, Patrick Fournier, Richard Greene, Yong Baek Kim, Hae-Young Kee, Steve Kivelson, Gilbert Lonzarich, Andrew Millis, Michael Norman, Subir Sachdev, Todadri Senthil and André-Marie Tremblay. I would like to thank Jacques Corbin for his help with several experiments. I would like to acknowledge the support of the Canadian Institute for Advanced Research and funding from NSERC, FQRNT, CFI and a Canada Research Chair.